\documentclass[sigconf]{acmart} %ACM sighit

\usepackage{flushend}

\makeatletter
\renewcommand\@formatdoi[1]{\ignorespaces}
\makeatother

\renewcommand\footnotetextcopyrightpermission[1]{} % removes footnote with conference information in first column
   \graphicspath{{./Img/}}
   \DeclareGraphicsExtensions{.pdf,.jpeg,.png,.jpg}
\usepackage{amsmath}

\usepackage{subfiles}

\usepackage{soul}
\usepackage{hyperref}

\usepackage{tikz}
\usepackage{pgfplots}
\usetikzlibrary{matrix}
\usepgfplotslibrary{groupplots}

\newcommand{\oursubsubsection}[1]{
	\vspace{0.5em}\subsubsection{#1} ~\par\vspace{0.25em} % cheating line break
}

\usepackage{float}
\floatstyle{boxed}
\restylefloat{table}
\usepackage{stfloats}

\hypersetup{
    colorlinks,
    linkcolor={red!50!black},
    citecolor={blue!50!black},
    urlcolor={blue!80!black}
}

\usepackage{enumitem}

\makeatletter
\newif\if@anonymize

\@anonymizefalse  % Uncomment to show text

\if@anonymize
  \newcommand{\highlight@DoHighlight}{
    \fill [outer sep = -15pt, inner sep = 0pt, color=lightgray]
          ($(begin highlight)+(0,8pt)$) rectangle ($(end highlight)+(0,-3pt)$) ;
  }

  \newcommand{\highlight@BeginHighlight}{
    \coordinate (begin highlight) at (0,0) ;
  }

  \newcommand{\highlight@EndHighlight}{
    \coordinate (end highlight) at (0,0) ;
  }

  \newdimen\highlight@previous
  \newdimen\highlight@current
  \newlength{\item@width}

  \DeclareRobustCommand*\anonymize{%
    \SOUL@setup
    \def\SOUL@preamble{%
      \begin{tikzpicture}[overlay, remember picture]
        \highlight@BeginHighlight
        \highlight@EndHighlight
      \end{tikzpicture}%
    }%
    \def\SOUL@postamble{%
      \begin{tikzpicture}[overlay, remember picture]
        \highlight@EndHighlight
        \highlight@DoHighlight
      \end{tikzpicture}%
    }%
    \def\SOUL@everyhyphen{%
      \discretionary{%
        \SOUL@setkern\SOUL@hyphkern
        \SOUL@sethyphenchar
        \tikz[overlay, remember picture] \highlight@EndHighlight ;%
      }{%
      }{%
        \SOUL@setkern\SOUL@charkern
      }%
    }%
    \def\SOUL@everyexhyphen##1{%
      \SOUL@setkern\SOUL@hyphkern
      \settowidth{\item@width}{##1}%
      \makebox[\item@width]{}%
      \discretionary{%
        \tikz[overlay, remember picture] \highlight@EndHighlight ;%
      }{%
      }{%
        \SOUL@setkern\SOUL@charkern
      }%
    }%
    \def\SOUL@everysyllable{%
      \begin{tikzpicture}[overlay, remember picture]
        \path let \p0 = (begin highlight), \p1 = (0,0) in \pgfextra
          \global\highlight@previous=\y0
          \global\highlight@current =\y1
        \endpgfextra (0,0) ;
        \ifdim\highlight@current < \highlight@previous
          \highlight@DoHighlight
          \highlight@BeginHighlight
        \fi
      \end{tikzpicture}%
      \settowidth{\item@width}{\the\SOUL@syllable}%
      \makebox[\item@width]{}%
      \tikz[overlay, remember picture] \highlight@EndHighlight ;%
    }%
    \SOUL@
  }
\else
  \newcommand{\anonymize}[1]{#1}
\fi
\makeatother

\acmConference[BER]{BER Workshop at HAI 2017 conference}{October 2017}{Bielefeld, Germany} 
\acmYear{2017}
\copyrightyear{2017}

\newcommand\blfootnote[1]{%
  \begingroup
  \renewcommand\thefootnote{}\footnote{#1}%
  \addtocounter{footnote}{-1}%
  \endgroup
}

\begin{document}

\title{Multimodal Observation and Interpretation of Subjects Engaged in Problem Solving}

\author{Thomas Guntz}
\affiliation{\institution{Univ. Grenoble Alpes, Inria, CNRS, Grenoble INP\textsuperscript{*}, LIG, F-38000 Grenoble, France}}
\email{Thomas.Guntz@inria.fr}

\author{Raffaella Balzarini}
\affiliation{\institution{Univ. Grenoble Alpes, Inria, CNRS, Grenoble INP\textsuperscript{*}, LIG, F-38000 Grenoble, France}}
\email{Raffaella.Balzarini@inria.fr}

\author{Dominique Vaufreydaz\vspace*{-1em}}
\affiliation{\institution{Univ. Grenoble Alpes, Inria, CNRS, Grenoble INP\textsuperscript{*}, LIG, F-38000 Grenoble, France}}
\email{Dominique.Vaufreydaz@inria.fr}

\author{James Crowley\vspace*{-1em}}
\affiliation{\institution{Univ. Grenoble Alpes, Inria, CNRS, Grenoble INP\textsuperscript{*}, LIG, F-38000 Grenoble, France}}
\email{James.Crowley@inria.fr}

%\author{\IEEEauthorblockN{Thomas Guntz\IEEEauthorrefmark{1},
%Raffaella Balzarini\IEEEauthorrefmark{1},
%Dominique Vaufreydaz\IEEEauthorrefmark{1}, and 
%James L. Crowley\IEEEauthorrefmark{1} 
%\IEEEauthorblockA{\IEEEauthorrefmark{1}Univ. Grenoble Alpes, Inria, CNRS, Grenoble INP, LIG, F-38000 Grenoble, France\\ Email: firstname.lastname@inria.fr}}}
%  This is the official signature to use. 
%\author{\IEEEauthorblockN{
%Authors\IEEEauthorrefmark{1}
%\IEEEauthorblockA{\IEEEauthorrefmark{1}Institution}}}


% \thispagestyle{plain}
% \pagestyle{plain}

\newcommand{\citehere}{ \textcolor{red}{\textit{citation needed}} }


%In this paper we present the first results of a pilot experiment in the capture and interpretation of multi-modal signals of human experts engaged in solving challenging problems.  
%We observed the eye gaze, fixation, body posture, facial expressions and cardiac signals for 21 chess experts. 
%We explore how such recordings can be used to estimate a subjects awareness of the current situation, and to predict ability to respond effectively to challenging situations. 
%Our goal in is to investigate the extent to which observations of eye-gaze, posture, emotion and other physiological signals can be used to model the cognitive state of subjects, and how explore the integration of multiple sensor modalities to improve the reliability of detection of human displays of awareness and emotion.
\begin{abstract}
In this paper we present the first results of a pilot experiment in the capture and interpretation of multimodal signals of human experts engaged in solving challenging chess problems.  
Our goal is to investigate the extent to which observations of eye-gaze, posture, emotion and other physiological signals can be used to model the cognitive state of subjects, and to explore the integration of multiple sensor modalities to improve the reliability of detection of human displays of awareness and emotion.
%We observed the eye gaze, fixations, body postures and facial expressions for chess players engaged in problems of increasing difficulty. 
%Primary results show that eye-gaze, body posture and emotions are good features
%to consider for determining expertise.
We observed chess players engaged in problems of increasing difficulty while recording their behavior.
Such recordings can be used to estimate a participant's awareness of the current situation and to predict ability to respond effectively to challenging situations. 
Results show that a multimodal approach is more accurate than a unimodal one.
By combining body posture, visual attention and emotion, the multimodal approach can reach up to 93\% of accuracy when determining player's chess expertise while unimodal approach reaches 86\%. 
Finally this experiment validates the use of our equipment as a general and reproducible tool for the study of participants engaged in screen-based interaction and/or problem solving.
% * <guntho.neo@gmail.com> 2017-10-11T09:13:42.464Z:
%
% ^.
\end{abstract} 

\keywords{\blfootnote{\textsuperscript{*} Institute of Engineering Univ. Grenoble Alpes}Chess Problem Solving, Eye Tracking, Multimodal Perception, Affective Computing}

\maketitle

\floatstyle{plain}
\restylefloat{table}
\begin{figure*}[ht] 
\centering
\includegraphics[width=1\linewidth]{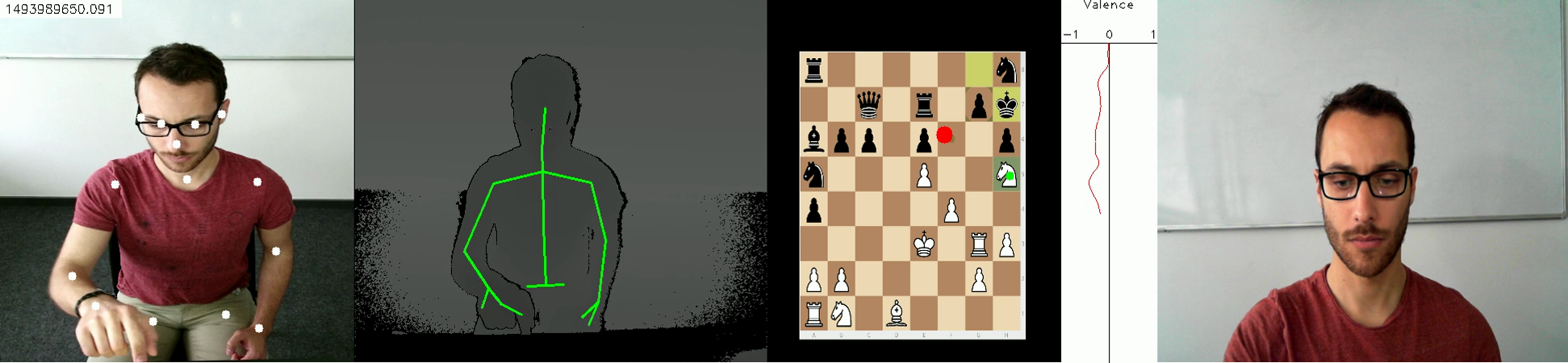}
%\vspace{-0.5cm}
\caption{Multimodal view of gathered data. Left to right: RGB (with body joints) and depth view from Kinect 2 sensors, screen record of chess task (red point is current position of gaze, green point is position of last mouse click), plot of current level of positive emotion expression (valence) and frontal view of face from webcam sensor.}
\label{fig:data}
%\vspace{-1cm}
\end{figure*}

% \fi

%\IEEEpeerreviewmaketitle

\vspace{0.5em}
\section{Introduction}

Commercially available sensing technologies are increasingly able to capture and interpret human displays of emotion and awareness through non-verbal channels. 
However, such sensing technologies tend to be sensitive to environmental conditions (\textbf{e.g.} noise, light exposure or occlusion), producing intermittent and unreliable information.
Techniques for combining multiple modalities to improve the precision and reliability of modeling of awareness and emotion are an open research problem.
Only few researches have been conducted so far on how such signals can be used to inform a system about cognitive processes such as situation awareness, understanding or engagement.
For instance, some researches showed that mental states can be inferred from facial expressions and gestures (from head and body) \cite{el2005real,baltruvsaitis2011real}.
%Little is known about how such signals can be used to inform a system about cognitive processes such as situation awareness, understanding, engagement, emotion or distress. 

Willing to increase focus on this area of research, we have constructed an instrument for the capture and interpretation of multimodal signals of humans engaged in solving challenging problems. 
Our instrument, shown in figure \ref{fig:setup}, captures eye gaze, fixations, body postures and facial expressions signals from humans engaged in interactive tasks on a touch screen.
As a pilot study, we have observed these signals for players engaged in solving chess problems. 
Recordings are used to estimate subjects' understanding of the current situation and their ability to respond effectively to challenging tasks.
Our initial research question for this experiment was:
\begin{itemize}[leftmargin=*]
\item \textit{Can our experimental set up be used to capture reliable recordings for such study?}
\end{itemize}
If successful, this should allow us to a second research question:
\begin{itemize}[leftmargin=*]
\item \textit{Can we detect when chess players are challenged beyond their abilities from such measurements?}
\end{itemize}
\par
% \vspace{-0.4cm} 

In this article, section 2 discusses current methods for capture and interpretation of physiological signs of emotion and awareness.
This lays the ground for the design of our experimental setup presented in section 3.
% We then discuss the problem of alignment of multiple modalities. 
Section 4 presents the results from our pilot experiment that was undertaken to validate our installation and evaluate the effectiveness of our approach. 
We conclude with a discussion on limitations and further directions to be explored in section 5.
%\hfill May 5, 2017

\section{State Of The Art}
Humans display awareness and emotions through a variety of non-verbal channels.
It is increasingly possible to record and interpret information from such channels. 
Thank to progress in related research, notably recently using Deep Learning approaches \cite{Morency2016,cao2017realtime,simon2017hand,wei2016cpm}, publicly available efficient software can be used to detect and track face orientation using commonly available web cameras.
Concentration can be inferred from changes in pupil size  \cite{Kanneman11}. 
Measurement of physiological signs of emotion can be done by detection of Facial Action Units  \cite{ekman1969nonverbal} from both sustained and instantaneous displays (micro-expressions). 
Heart rate can be measured from the Blood Volume Pulse as observed from facial skin color  \cite{poh2011advancements}.
Body posture and gesture can be obtained from low-cost RGB sensors with depth information (RGB+D)  \cite{shotton2013real}.
Awareness and attention can be inferred from eye-gaze (scan path) and fixation using eye-tracking glasses as well as remote eye tracking devices  \cite{stiefelhagen1997model}. 
This can be directly used to reveal cognitive processes indicative of expertise  \cite{charness2001perceptual} or situation awareness in human-computer interaction (HCI) systems \cite{Paletta2017}. 
However, the information provided by each of these modalities tends to be intermittent, and thus unreliable. 
Most investigators seek to combine multiple modalities to improve both reliability and stability  \cite{giraud2013multimodal,abadi2013multimodal}.

Chess analysis has long been used in Cognitive Science to understand attention and to develop models for task solving. 
In their study \cite{charness2001perceptual,reingold2005perception}, Charness \textit{et al} showed that when engaging in competitive game, chess players display engagement and awareness of the game situation with eye-gaze and fixation. %and other signs of attention.
This suggests that the mental models used by players can be at least partially determined from eye gaze, fixation and physiological response. 
The ability to detect and observe such models during game play can provide new understanding of the cognitive processes that underlay human interaction.
Experiments described in this article are the preamble to more advanced research on this topic. 
 
\section{Experiments}
\label{chap:Methods_Analysis}
As a pilot study, chess players were asked to solve chess tasks within a fixed, but unknown, time frame. 
We recorded eye gaze, facial expressions, body postures and physiological reactions of the players as they solved problems of increasing difficulty.
The main purpose is to observe changes in their reactions when presented tasks are beyond their level.

\subsection{Materials and Participants}

\begin{figure}[!t]
\centering
\includegraphics[width=\linewidth, angle=-90]{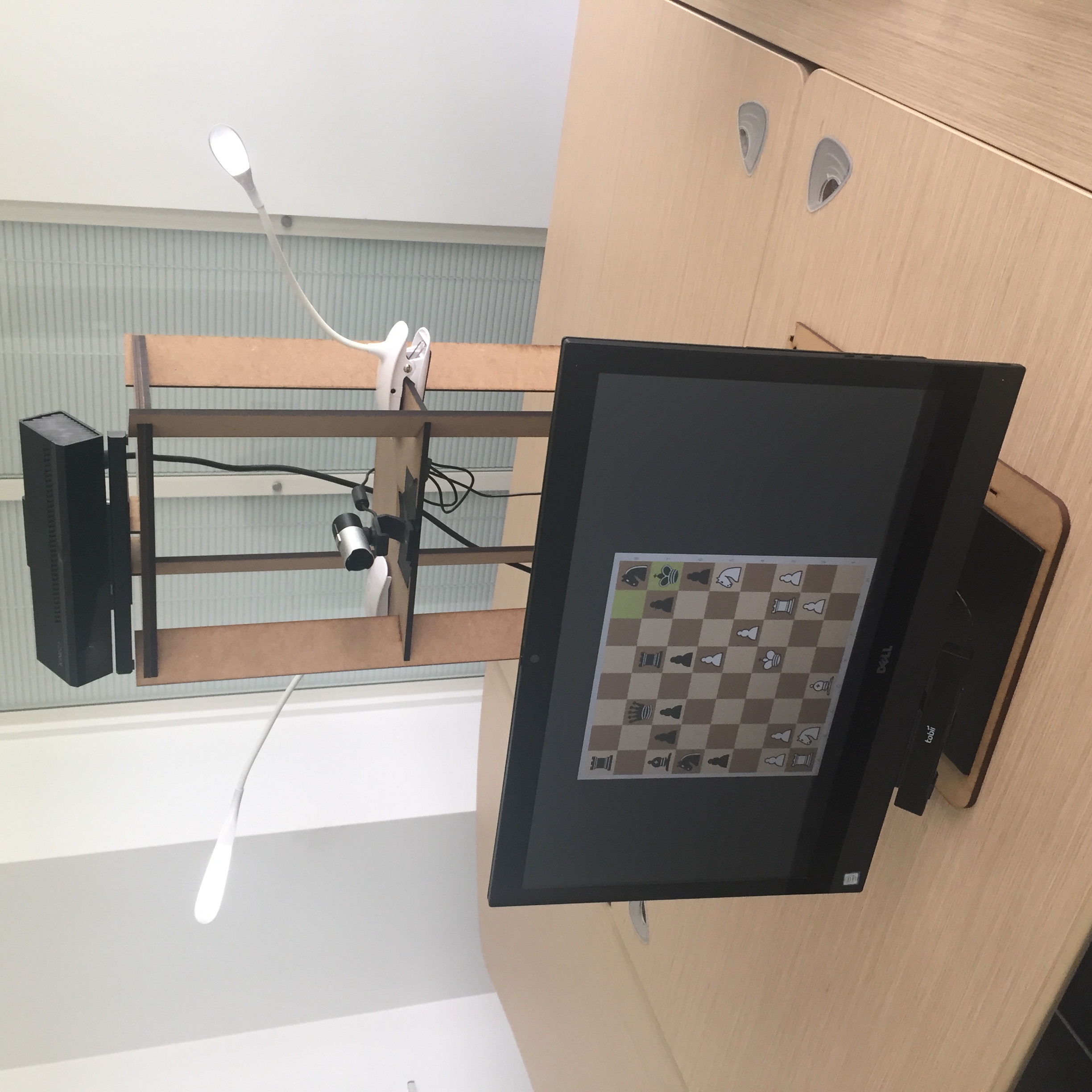}
\caption{The experimentation equipment used for data collection. On top, a Kinect2 device looking down at the player. In the middle, a webcam to capture the face. At bottom, the touch screen equipped with an eye-tracker presenting the chess game. These views are respectively at left, right and center of figure \protect{\ref{fig:data}}. The wooden structure is rigid to fix position and orientation of all sensors. The lighting conditions are controlled by 2 USB LED lamps on the sides.}
\label{fig:setup}
\vspace{-0.5cm}
\end{figure}

% \vfill\break
\oursubsubsection{Experimental setup}

Figure~\ref{fig:setup} presents the recording setup for our experiment.
This setup is a derivative version of the one we use to record children during storytelling sessions \cite{portaz:hal-01595775}.
As seen, it is composed of several hardware elements: a 23.8" Touch-Screen computer, a Kinect 2.0 mounted 35cm above the screen focusing on the chess player, a 1080p Webcam for a frontal view, a Tobii Eye-Tracking bar (Pro X2-60 screen-based) and two adjustable USB-LED for lighting condition control.
The use of the Touch-Screen during the entire experiment was chosen to provide a gesture-based play resembling play with a physical board.
A wooden super-structure is used to rigidly mount the measuring equipment with respect to the screen in order to assure identical sensor placement and orientation for all recordings. 
This structure have been made using a laser cutter.

Several software systems were used for recording and/or analyzing data.
The Lichess Web Platform\footnote{~\url{https://en.lichess.org/} (last seen 09/2017)}~ % Lichess webmaster gave their approval to use their platform for this scientific experiment.%} 
serves for playing and recording games.
Two commercial software provide both online and offline information: Tobii Studio 3.4.7 for acquisition and analyze of eye-gaze; Noldus FaceReader 7.0 for emotion detection.
Body postures information were given by two different means: by the Kinect 2.0 SDK and by using our enhanced version of the Realtime Multi-Person Pose Estimation software  \cite{cao2017realtime}. 
Considering the state-of-the-art results of the second software, we decided to keep only this one for this experiment.
%Body postures were given both by the Kinect 2.0 SDK and using an enhanced version of the Realtime Multi-Person Pose Estimation software  \cite{cao2017realtime}.
%\todo{The last one was more accurate in our case. (belongs to the discussion and need to be explained)}
During the study, data were recorded from all sensors (Kinect 2, Webcam, Screen capture, user clicks, Tobii-Bar) using the RGBD Sync SDK\footnote{~\url{https://github.com/Vaufreyd/RGBDSyncSDK} (last seen 09/2017)} from the MobilRGBD project \cite{vaufreydaz:hal-01095667}.
This framework permits to read recorded and further computed data (gaze fixation, emotion detection, body skeleton position, etc.) for synchronous analysis by associating a timestamp with a millisecond precision to each recorded frame.
The same framework can read, analyze and display the same way all gathered or computed data.
An example is presented on figure~\ref{fig:data} where most of the data are depicted.

\oursubsubsection{Participants}
\label{sec:participants}

An announcement for our experiment with an invitation to participate was communicated to chess clubs, on the local university campus and within the greater metropolitan area.
We received a positive response from the president of one of the top metropolitan area chess clubs, and 21 members volunteered to participate in our pilot experiment.  
Unfortunately, of these initial 21 participants, 7 recordings were not usable due to poor eye-tracking results and have not been included in our analysis.
Indeed, these participants, while reflecting about the game, held their hand above the eye-tracker and disrupted its processing. 

% All participants were asked to solve 15 chess tasks while being recorded.
The 14 remaining chess players in our study were 7 experts and 7 intermediates level players (20-45 years, 1 female, age: $M=31.71; SD=7.57$).
Expert players were all active players and with \textit{Elo} ~ratings\footnote{~The \textit{Elo} system is a method to calculate rating for players based on tournament performance. Ratings vary between 0 and approximately 2850. \url{https://en.wikipedia.org/wiki/Elo_rating_system} (last seen 09/2017)}
ranged from 1759 to 2150 ($M = 1950; SD = 130$). %Typically, experts are rated between 2000 and 2200 points; masters are between 2200 and 2399; grandmasters are above 2500.
For the intermediate players, the \textit{Elo} ratings ranged from 1399 to 1513 ($M = 1415; SD = 43$) and 6 among them were casual players who were not currently playing in club.
We can also give some statistics on the recorded session: 
the average recording time per participant is 14:58 minutes ($MIN=4$:$54$, $MAX=23$:$40$, $SD=5$:$26$) and the average compressed size of gathered data is $56.12$ GiB per session.

\subsection{Methods}

\oursubsubsection{Chess Tasks}

The goal of this experiment was to engage participants into a cognitive process while observing their physiological reactions.
Thirteen chess tasks were elaborated by our team in coordination with the president of the chess club. % (Elo=1466). 
Two kinds of task were selected: \textit{chess openings tasks}, where only 3 to 5 moves were played from the original state; and \textit{N-Check-Mate tasks}, where 1 to 6 moves were required to check-mate the opponent (and finish the game).

\textbf{Openings.}
Skilled players are familiar with most of the chess openings and play them intuitively.
Intuitive play does not generally require cognitive engagement for reasoning. 
An important challenge is to detect when a player passes from intuitive reaction to a known opening, to challenging situations. 
Thus, two uncommon openings were selected to this end: a King's Gambit (3 moves from the initial state) and a Custom Advanced Variation of the Caro-Kann Defense (6 moves from initial state). 
The goal here is to pull participants out from their comfort zone as much as possible to evoke emotions and physiological reactions.
Openings correspond to task number 1 and 2.

\textbf{N-Check-Mate.}
Eleven end game tasks were defined. 
These are similar to the daily chess puzzles that can be found in magazines or on chess websites.
Each of these tasks was designed to check-mate the opponent in a number of predefined moves ranging from 1 to 6. 
Tasks requesting 1 to 3 moves are viewed as easy task whereas 4 to 6 moves tasks require more chess reasoning abilities, etc.
%Distribution among tasks differ according to the difficulty as seen in table~\ref{tab:chess_tasks}. 
Distribution among the 11 tasks differs according to their number of required move and thus to their difficulty:
4 tasks with one move, 4 tasks with two and three moves (2 of each) and 3 tasks with four, five and six moves (1 of each).
End games were presented to participants in this order of increasing difficulty while alternating the played color (white/black) between each task.

\oursubsubsection{Procedure}

Participants were tested individually in sessions lasting approximately 45 minutes. 
Each participant was asked to solve the 13 chess tasks and their behaviors were observed and recorded.
To avoid biased behavior, no information was given about the recording equipment.
Nevertheless, it was necessary to reveal the presence of the eye-tracker bar to participants in order perform a calibration step. 
After providing informed consent, the Lichess web platform was presented and participants could play a chess game against a weak opponent (\textit{Stockfish}\footnote{~\textit{Stockfish} is an open-source game engine used in many chess software, including Lichess. \url{https://en.wikipedia.org/wiki/Stockfish_(chess)} (last seen 09/2017). } algorithm level 1: lowest level) to gain familiarity with the computer interface.
No recording was made during this first game.\par

Once familiar and comfortable with the platform, the eye-tracking calibration was performed using Tobii Studio software, in which subjects were instructed to sit between 60 and 80cm from the computer screen and to follow a 9-point calibration grid. 
Participants were requested to avoid large head movement in order to assure good eye-tracking quality.
Aside from this distance, no other constraints were instructed to participants.

Each task to solve was individually presented, starting with the openings, followed by the N-Check-Mate tasks.
Participants were instructed to solve the task by either playing a few moves from the opening or to check mate the opponent (played by \textit{Stockfish} algorithm level 8: the highest level) in the required number of moves.
The number of moves needed for the N-Check-Mate tasks was communicated to the subject. 
A time frame was imposed for each task.
The exact time frame was not announced to the participant, they only knew that they have a couple of minutes to solve the task.
This time constraint ranges from 2 minutes for the openings and the easiest N-Check-Mate tasks (1-2 moves) to 5 minutes for the hardest ones (4-5-6 moves). 
An announcement was made when only one minute was remaining to solve the task.
If the participant could not solve the task within the time frame, the task was considered as failed and the participant proceeded to the next task.
The experiment is considered finished once all tasks were presented to the participant.

\vfill\break
\subsection{Analysis}
\oursubsubsection{Eye-Gaze}

Eye movement is highly correlated with focus of attention and engaged cognitive processes \cite{holmqvist2011eye} in problem solving and human-computer interaction \cite{poole2006eye}. 
Other studies  \cite{charness2001perceptual,reingold2005perception} show that expertise estimation for chess players can be performed using several eye-tracking metrics such as fixation duration or visit count.
In this case, gaze information can be useful to determine information such as:
\begin{enumerate}[leftmargin=*]
  \item \textit{What pieces received the most focus of attention from participants?}
  \item \textit{Do participants who succeed to complete a task share the same scan path?}
  \item \textit{Is there significant difference in gaze movements between novices and experts?}
\end{enumerate}
To reach these aims, Areas Of Interests (AOIs) were manually defined for every task.
An AOI can be a key piece for the current task (\textbf{e.g.} a piece used to check-mate the opponent), the opponent king, destinations cases where pieces have to be moved, etc.
Afterward, we compute statistics for every AOI of each task.
Among possible metrics, results depicted in this article are based on \textit{Fixation Duration}, \textit{Fixation Count} and \textit{Visit Count}.

Interpretation for these metrics differs according to the task domain. 
For example, in the domain of web usability, Ehmke et al  \cite{ehmke2007identifying} would interpret long fixation duration on AOI as a difficulty to extract or interpret information from an element. 
In the field of chess, Reingold and Charness  \cite{reingold2005perception,charness2001perceptual} found significant differences in fixation duration between experts and novices.
%\par For this study, a particular focus has been done on proportion of fixations for participants. 

\oursubsubsection{Facial emotions}

Micro-expressions, as defined by Ekman and Fiesen  \cite{ekman1969nonverbal} in 1969, are quick facial expressions of emotions that could last up to half a second. 
These involuntary expressions can provide information about cognitive state of chess players. 
In our pilot study, the Noldus FaceReader software  \cite{facereader_presentation} has been used to classify players' emotions in the form of six universal states proposed by Ekman: happiness, sadness, anger, fear, disgust and surprise (plus one neutral state).  
These emotional states are commonly defined as regions in a two-dimensional space whose axes are valence and arousal.
Valence is commonly taken as an indication of pleasure, whereas arousal describes the degree to which the subject is calm or excited. \par

In practice, the FaceReader software analyses video by first applying a face detector to identify a unique face followed by a detection of 20 Facial Action Units  \cite{ekman1969nonverbal}. 
Each action unit is assigned a score between 0 and 1 and these are used to determine the state label for emotion. Valance and arousal can be then computed as: 
\begin{itemize}[leftmargin=*]
\item \textbf{Valence}: intensity of positive emotions (\textit{Happy}) minus intensity of negatives emotions (\textit{sadness}, \textit{anger}, \textit{fear} and \textit{disgust});
\item \textbf{Arousal}: computed accordingly to activation intensities of the 20 Action Units.
\end{itemize}
FaceReader was tested on two different datasets: the Radboud Faces Database  \cite{radboud} containing 59 different models and the Karolinska Directed Emotional Faces  \cite{karolinska} which regroups 70 individuals. 
Both dataset display 7 different emotional expressions (plus neutral) on different angles. 
FaceReader algorithm correclty classified 90\% of the 1197 images from Radboud Face Database  \cite{facereader_radboud} and 89\% of the Karolinska Dataset (4900 images)  \cite{facereader_presentation}.

\oursubsubsection{Body Posture}

Body posture is  a rich communication channel for human to human interaction with important potential for human computer interaction  \cite{anzalone2015evaluating}.
Studies have shown that self-touching behavior is correlated with negative affect as well as frustration in problem solving  \cite{harrigan1985self}.
Thus, we have investigated a number of indicators for stress from body posture: 
\begin{itemize}[leftmargin=*]
\item \textbf{Body Agitation}: how many joints are varying along $x$, $y$ and $z$ axis;
\item \textbf{Body Volume}: space occupied by the 3D bounding box built around joints (see  \cite{johal2015cognitive});
\item \textbf{Self-Touching}: collisions between wrist-elbow segments and the head (see  \cite{aigrain2016multimodal}).
\end{itemize}

These signals are computed from the RGBD streams recorded by the Kinect 2 where a list of body joints is extracted by means of our variant of a body pose detection algorithm \cite{cao2017realtime}.
These joints are computed on the RGB streams and projected back to Depth data.
Thus, a 3D skeleton of the chess player is reconstructed and can be used as input to compute previous metrics. 
As one can see on figures \ref{fig:data} at left, from the point of view of the Kinect 2 in our setup (see figure \ref{fig:setup}), the skeleton information is limited to the upper part of the body, from hips to head.

%\begin{table}[h!]
%   \centering   
%   \caption{Features of interest}
%   \label{tab:features}
%   \begin{tabular}{|c|c|c|}
%   \hline
%   Features & Modality & Sensor \\
%   \hline
%   Fixation Duration &   &   \\
%   Fixation Count 	 & Eye-Gaze & Tobii Bar \\
%   Visit Count 	 &   &   \\
%   \hline
%   \begin{tabular}{@{}c@{}}Valence \\ Arousal\end{tabular}& Emotion & Webcam \\
%   \hline
%   Agitation &  & \\
%   Volume & Body & Kinect \\
%   Self-Touch &  & \\
%   \hline
%   \end{tabular}
%\end{table}

\section{Results}

\label{chap:results}
Synchronous data for every feature have been extracted from all sensors.
Several tasks, like regression over \textit{Elo} ratings or over the time needed to perform a task, could be addressed using these data.
Among them, we chose to analyze a classification problem that can be interpreted by a human:
\begin{itemize}[leftmargin=*]
\item \textit{Is it possible, by the use of gaze, body and/or facial emotion features, to detect if a chess player is an expert or not?}
\end{itemize}
This problem is used as example to obtain a first validation of our data relevancy.
It is correlated with whether a chess player is challenging beyond his abilities. 

%Primary results have been extracted from these experiments to compare unimodal and multimodal analysis of chess players' expertise.
This section presents unimodal and multimodal analysis of extracted features to determine chess expertise of players.
%This section presents how using gaze, body and/or facial emotion features, one can detect if the player is an expert or not.
%Data of every sensors have been extracted from these experiments and are first analyzed individually (unimodal approach) before combining them (multimodal approach). 

\subsection{Unimodal analysis}

\oursubsubsection{Eye-Gaze}

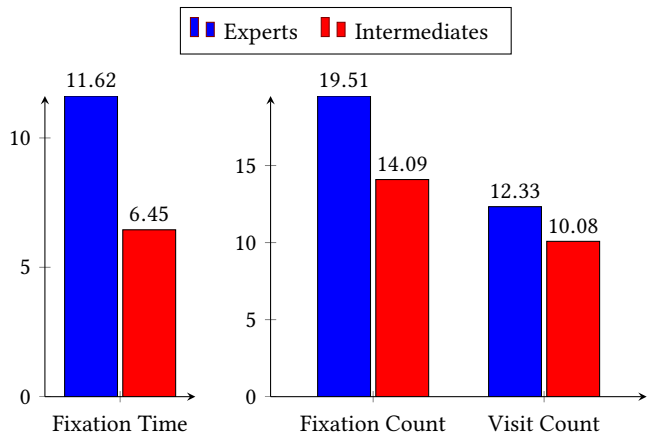
\begin{figure}[!ht]
\begin{center}
\begin{tikzpicture}
\pgfplotsset{
    scale only axis,
    compat=1.3
}
\begin{groupplot}[group style={group name=myplot, group size=2 by 1},ymin=0,height=5cm,width=5cm]
  \nextgroupplot[
        ybar,
        bar width=20pt,
        ytick={0,5,10,15,20},
        xtick=data,
        axis x line=bottom,
        axis y line=left,
        enlarge x limits=0.3,
        symbolic x coords={Fixation Time},
        xticklabel style={anchor=base,yshift=-\baselineskip},
        height=4cm,width=2cm,
        nodes near coords
      ]
        \addplot[fill=blue] coordinates {
        %expert
          (Fixation Time, 11.62)
          };\label{plots:plot1}

          \addplot[fill=red] coordinates {
          %nonexpert
          (Fixation Time,6.45)
        };\label{plots:plot2}
        
  \nextgroupplot[
        ybar,
      axis y line=left,
        bar width=20pt,
      ytick={0,5,10,15,20},
        xtick=data,
        axis x line=bottom,
        enlarge x limits=0.6,
        symbolic x coords={Fixation Count, Visit Count},
        xticklabel style={anchor=base,yshift=-\baselineskip},
        height=4cm,width=5cm,
        nodes near coords
      ]
        \addplot[fill=blue] coordinates {
        %expert
          (Fixation Count, 19.51)
          (Visit Count, 12.33)
          };

          \addplot[fill=red] coordinates {
          %nonexpert
          (Fixation Count, 14.09)
          (Visit Count, 10.08)
        };
%		Expert	Intermediates
%Fixation_Duration_Mean	Fixation_Duration_Mean	11.6256678352	6.4471789718
%Fixation_Count_Mean	Fixation_Count_Mean	19.5194805195	14.0909090909
%Visit_Count_Mean	Visit_Count_Mean	12.3376623377	10.0779220779
  \end{groupplot}
  
% legend
\path (myplot c1r1.north west|-current bounding box.north)--
      coordinate(legendpos)
      (myplot c2r1.north east|-current bounding box.north);
\matrix[
    matrix of nodes,
    anchor=south,
    draw,
    inner sep=0.2em,
    draw
  ]at([yshift=1ex]legendpos)
  {
    \ref{plots:plot1}& Experts&[5pt]
    \ref{plots:plot2}& Intermediates&[5pt]\\};
\end{tikzpicture}
\end{center}
\caption{Eye-gaze histograms. Left: Percentage of fixation (in time). Right: number of fixations and number of visits.}
\label{fig:results:gaze}
\end{figure}

Two AOIs were defined for each task: one AOI is centered on the very first piece to move in the optimal sequence to successfully achieve the check-mate; and the second one on the destination square where this piece has to be moved.
Fixations and visits information of every task are gathered for all participants and results are presented in Figure~\ref{fig:results:gaze}.

As can be clearly seen in this figure, experts have longer and more fixations than intermediates on relevant pieces.
Same result is observed for visit count.
Similar results can be found in literature  \cite{charness2001perceptual}.
These results are explained by the expert's skill encoding capacity that enables them to quickly focus their attention on relevant piece by a better pattern matching ability.

More work has to be done on eye-gaze such as analyzing and comparing the scan path order of participants, measuring how fast are participants to identify relevant pieces or analyzing fixation on empty squares.

\oursubsubsection{Emotions}

The increasing difficulty in the non-interrupting tasks has caused our participants to express more observable emotions across the experiment.
Emotions in a long-task experiment are expressed as peaks in the two-dimensional space (valence, arousal). 
Thus, standard statistics tend to shrink toward zero as the record becomes longer. 

Other approaches should be considered to visualize emotion expressions. 
One possibility is to consider the number of changes of emotions having the highest intensity (i.e. the current detected facial emotion).
As emotion intensities are based on facial unit detection, changes in the main emotion denote underlying changes in facial expression.
The result metric is shown on the graph presented in figure~\ref{fig:results:analysis}.

\begin{figure}[!ht]
\parbox{\linewidth}{
\centering
\begin{tikzpicture}
\pgfplotsset{compat=1.14}{
    scale only axis,
    width=7cm,
    height=5cm, 
    compat=1.3
}
\begin{axis}[
    xlabel={},
    ylabel={},
    xmin=1, xmax=13, xtick={1,2,3,4,5,6,7,8,9,10,11,12,13},
    ymin=0, ymax=15,
    xlabel={Task Number},
    ylabel={Average count of main emotion change},
    legend pos=north west,
    ymajorgrids=true,
    grid style=dashed,
]
\addplot[
	%expert
    color=blue,
    mark=square,
    ]
    coordinates {
    (1,1.14)
    (2,5.14)
    (3,1.0)
    (4,0.71)
    (5,0.14)
    (6,0.29)
    (7,0)
    (8,0.71)
    (9,2.0)
    (10,2.29)
    (11,4.29)
    (12,5.86)
    (13,9.14)
    };
    
\addplot[
	%nonexpert
    color=red,
    mark=square,
    ]
    coordinates {
    (1,2.57)
    (2,6.71)
    (3,3.86)
    (4,2.00)
    (5,0.86)
    (6,0.71)
    (7,2.43)
    (8,1.57)
    (9,11.29)
    (10,8.14)
    (11,12.14)
    (12,10.0)
    (13,8.29)
    };    
    
%		expert	nonexpert
%Total Sum - 1_Kings_Gambit		8	18
%Total Sum - 2_Caro_Kann		36	47
%Total Sum - 3_IV		7	27
%Total Sum - 4_IX		5	14
%Total Sum - 5_V		1	6
%Total Sum - 6_XVI		2	5
%Total Sum - 7_XVII		0	17
%Total Sum - 8_XXIX		5	11
%Total Sum - 9_XLIII		14	79
%Total Sum - 10_LI		16	57
%Total Sum - 11_LIX		30	85
%Total Sum - 12_LXVII		41	70
%Total Sum - 13_LXIX		64	58

%		expert	nonexpert
%Total Sum - 1_Kings_Gambit		1.1428571429	2.5714285714
%Total Sum - 2_Caro_Kann		5.1428571429	6.7142857143
%Total Sum - 3_IV		1	3.8571428571
%Total Sum - 4_IX		0.7142857143	2
%Total Sum - 5_V		0.1428571429	0.8571428571
%Total Sum - 6_XVI		0.2857142857	0.7142857143
%Total Sum - 7_XVII		0	2.4285714286
%Total Sum - 8_XXIX		0.7142857143	1.5714285714
%Total Sum - 9_XLIII		2	11.2857142857
%Total Sum - 10_LI		2.2857142857	8.1428571429
%Total Sum - 11_LIX		4.2857142857	12.1428571429
%Total Sum - 12_LXVII		5.8571428571	10
%Total Sum - 13_LXIX		9.1428571429	8.2857142857
\legend{experts, intermediates}
\end{axis}
\end{tikzpicture}
\caption{Average count of variation of main detected facial emotion in regard to the task (1-13). Tasks are ranging in an increasing difficulty order.}
\label{fig:results:analysis}
}
\parbox{\linewidth}{
\centering
\begin{tikzpicture}
\pgfplotsset{compat=1.14}{
    scale only axis,
    width=7cm,
    height=5cm, 
    compat=1.3
}
\begin{axis}[
    xlabel={},
    ylabel={},
    xmin=1, xmax=13,xtick={1,2,3,4,5,6,7,8,9,10,11,12,13},
    ymin=0, ymax=2.5,
    xlabel={Task Number},
    ylabel={Average count of self-touching},
    legend pos=north west,
    ymajorgrids=true,
    grid style=dashed,
]
\addplot[
    color=blue,
    mark=square,
    ]
    coordinates {
    %ADD The coordinates here for the first plot
    (1,0.57)
    (2,0.29)
    (3,0.14)
    (4,0.29)
    (5,0)
    (6,0.29)
    (7,0.14)
    (8,0)
    (9,0.43)
    (10,0.14)
    (11,0.7)
    (12,0.57)
    (13,1)
    };
    
\addplot[
    color=red,
    mark=square,
    ]
    coordinates {
    %ADD The coordinates here for the second plot
    (1,0.14)
    (2,0.86)
    (3,0.29)
    (4,0.57)
    (5,0.43)
    (6,0.29)
    (7,0.29)
    (8,0.14)
    (9,1.43)
    (10,1.43)
    (11,1.43)
    (12,2)
    (13,0.88)
    };    
\legend{experts, intermediates}
\end{axis}
\end{tikzpicture}
\caption{Average count of self-touching in regard to the task (1-13). Tasks are ranging in an increasing difficulty order.}
\label{fig:results:selftouch}
}
\end{figure}

It clearly appears that expression of emotions increase with the difficulty of the problem to solve.
For both player classes, there is a peak for the second task (i.e. our uncommon custom advanced variation of the Caro-Kann defense).
This opening was surprising for all participants, more than the King's
Gambit one (task 1).
No participant was familiar with this kind of opening.
Moreover, intermediates players present an emotional peak at task number 9, which is the first task to require more than 2 moves to check-mate the opponent, whereas expert's plot shape looks more like the beginning of an exponential curve. 
An interesting aspect of that plot is the final decrease of intermediate players after task 10, this could be interpreted as a sort of resignation, when players knew that tasks beyond of their skills and could not be resolved.
% Task 2's peak in emotions is caused by the uncommon opening that has been chosen for the task. 

These primary results suggest that situation understanding and expertise knowledge can be inferred from variation of facial emotions.
Although, more detailed analysis, such as activation of Action Units, derivative of emotions or detection if a micro expression occurs right after a move being played should be performed.

\oursubsubsection{Body Posture}

The increasing difficulty of the N-Check-Mate tasks is a stress factor that can be observable according to \cite{harrigan1985self}.
Using technique presented in  \cite{aigrain2016multimodal} to detect self-touching, we can observe how participants' body reacts to the increasing difficulty of tasks.

% \begin{figure}[!h]
% \begin{center}

% \end{figure}

%Expert	0.5714285714	0.2857142857	0.1428571429	0.2857142857	0	0.2857142857	0.1428571429	0	0.4285714286	0.1428571429	0.7142857143	0.5714285714	1
%NonExpert	0.1428571429	0.8571428571	0.2857142857	0.5714285714	0.4285714286	0.2857142857	0.2857142857	0.1428571429	1.4285714286	1.4285714286	1.4285714286	2	0.8571428571

%TO BE DELETED
%\begin{figure}[!h]
%\centering
%\includegraphics[width=0.95\linewidth]{results_selftouch}
%\caption{(TODO: put better Figure)Average count self-touching}
%\label{fig:results:selftouch}
%\end{figure}

The figure~\ref{fig:results:selftouch} presents statistics about self-touching.
Shapes of lines are very similar of what is observed for emotions (Figure~\ref{fig:results:analysis}).
The same conclusion can be drawn: the number of self-touches increases as tasks get harder and it reveals that this is a relevant feature to consider.
However, analysis of volume and agitation features did not reveal interesting information yet.
This can be explained either by the nature of the task or by the number of analyzed participants. 
More discussion of this experiment can be found in section~\ref{chap:Discussion}.

\subsection{Multimodal versus unimodal classification}

To demonstrate the potential benefit of a multimodal approach, a supervised machine learning algorithm has been used to quantify accuracy of different modalities for classification.
Only the data recorded for the 11 N-Check-Mate tasks are considered here.
Support Vector Machines (SVM) have been built for each modality and for each possible combination of modalities.\par
After computing statistical analysis (mean, variance, standard deviation) over our features, two approaches are compared: a task dependent approach on one hand and a skill-only dependent (All Task) on the other hand.
First approach considers statistical results for every participant and for every task.
That way, one input sample would be the instantiation of one participant for one particular task, given a total number of $14*11=154$ input samples.
Second approach takes the average over all tasks for each participant.
Input sample is reduced to participant with average statistics over tasks as features.

\begin{table*}[t!]
   \centering
   \caption{Best accuracy scores from cross-validation for SVMs. First line is Task Dependent approach, the number of sample $N$ is the number of participants (14) times the number of N-Check-Mate tasks (11). Second approach uses only average data of all task for every participant ($N$=14). Columns are the modality subset chosen to train the SVM (G: Gaze, B: Body, E: Emotions).}
   \label{tab:SVM_Accuracy}
   \resizebox{0.80\linewidth}{!}{%
   \begin{tabular}{|c|c|c|c|c|c|c|c|}
   \hline%\vspace{0.25em}
   ~ & G & B & E & G + B & G + E & B + E & G + B + E \\
   \hline
   Task Dependent ($N=154$)& $62\%$ & $58\%$ & $78\%$ & $58\%$ & $79\%$ & $79\%$ & $78\%$ \\
   \hline
   All Tasks ($N=14$) & $71\%$ & $79\%$ & $86\%$ & $71\%$ & $86\%$ & $93\%$ & $93\%$ \\
   \hline
   \end{tabular}%
   }
   \vspace{-0.2em}
\end{table*}

Stratified cross-validation procedure has been used on every SVM and for both approaches to compute their accuracy.
Results are shown in table~\ref{tab:SVM_Accuracy}. 
First observations of these results show that the task dependent approach presents a far less accuracy score than the second approach.
This could be explained by the variation in the length of recordings.
Indeed, some participants managed to give an answer in less than 10 seconds.
The second hypothesis shows good performance and validates one of our expectation that multimodal system could outperform unimodal ones.
Even if these scores are promising, further experiments with more involved participants have to be performed to confirm these primary results.

\section{Discussion}

\label{chap:Discussion}
% In human-human interaction, people use more than one modality to communicate. 
% HCI systems relying on more than one modality should perform better in human understanding. 
This research and primary results (see section \ref{chap:results}) show consistency results on unimodal features used to distinguish expert and intermediate chess players. 
When used together, body posture, visual attention and emotion provide better accuracy using a binary SVM classifier.
Although these results appear promising, they are only preliminary: the number of participants (14); the variation of recording duration (from seconds to a couple of minutes depending on the task and player’ expertise); and the tasks must all be expanded and developed.
% This ends to a record length ranging from 10 minutes for some experts players to 30 minutes for other participants.
Due to the size of our dataset, generalizing this preliminary results is not possible for the moment.
Further experiments must be conducted to validate them.

The conditions of the chess tasks should also draw attention. 
In the experimental configuration, chess players were facing a chess algorithm engine in tasks where they knew the existence of a winning sequence of moves.
Moreover, players are seating (see figure \ref{fig:data}), some clues like body agitation may provide less information than expected.
Participants may not be as engaged as they would have been in a real chess tournament facing a human opponent using an actual chess board.
In these particular situations, involving stakes for players, the physiological reactions and emotional expressions are more interesting to observe.

% However, the recording setup used for this study is not suited yet for a human-human chess game.\par
Nevertheless, these experiments reveal that valuable information can be observed from human attention and emotions to determine understanding, awareness and affective response to chess solving problems.
Another underlying result is the validation of our setup in monitoring chess players.
The problems encountered with the eye-tracker for 7 participants (see section \ref{sec:participants}) show that we must change its position to increase eye-tracking accuracy.

\section{Conclusion}
%We present through this first experiment how to integrate multiple sensors and how to perform unimodal and multimodal detection of awareness. \par
%Primary findings showing that multi-modal approach can improve detection of participants' awareness are quite interesting and promising.
%These results encourage us to repeat further experiment by increasing the number of participants.
%Another perspective is to perform more analysis on current signals: scan path from gaze, duration of self-touch or detection of emotional peak; and %explore new modalities: audio procedural speech or cardiac pulse rate.
%\par
%The anonymous collected database will be made publicly available.
This paper presents results from initial experiments with the capture and interpretation of multi-modal signals of 14 chess players engaged in solving 13 challenging chess tasks.
Results show that eye-gaze, body posture and emotions are good features to consider. 
Support Vector Machine classifiers trained with cross-fold validation revealed that combining several modalities could give better performances (93\% of accuracy) than using a unimodal approach (86\%).
These results encourage us to perform further experiments by increasing the number of participants and integrating more modalities (audio procedural speech, heart rate etc.).

Our equipment is based on off-the-shelf commercially available components as well as open source programs and thus can be easily replicated.
In addition to providing a tool for studies of participants engaged in problem solving, this equipment can provide a general tool that can be used to study the effectiveness of affective agents in engaging users and evoking emotions.


%\ifCLASSOPTIONcompsoc
  \section*{Acknowledgments}
%\else
%  \section*{Funding and Acknowledgment}
%\fi

\anonymize{This research has been be funded by the French ANR project CEEGE (ANR-15-CE23-0005), and was made possible by the use of equipment provided by ANR Equipement for Excellence Amiqual4Home (ANR-11-EQPX-0002). Access to the facility of the MSH-Alpes SCREEN platform for conducting the research is gratefully acknowledged.

We are grateful to all of the volunteers who generously gave their time to participate in this study and to Lichess webmasters for their help and approval to use their platform for this scientific experience.
We would like to thank Isabelle Billard, current chairman of the chess club of Grenoble ''\textit{L'\'Echiquier Grenoblois}`` and all members who participated actively in our experiments.}

%This research has been be funded by the French ANR project CEEGE (ANR-15-CE23-0005), and was made possible by the use of equipment provided by ANR Equipement for Excellence Amiqual4Home (ANR-11-EQPX-0002).\par
%We are grateful to all of the volunteers who generously gave their time to participate in this study.
%We would like to thank Isabelle Billard, current chairman of the chess club of Grenoble ''L'Echiquier Grenoblois`` and all members who participated actively in our experiments.

\vfill\break
\bibliographystyle{IEEEtran}
\bibliography{Biblio}

\begin{thebibliography}{10}
\providecommand{\url}[1]{#1}
\csname url@samestyle\endcsname
\providecommand{\newblock}{\relax}
\providecommand{\bibinfo}[2]{#2}
\providecommand{\BIBentrySTDinterwordspacing}{\spaceskip=0pt\relax}
\providecommand{\BIBentryALTinterwordstretchfactor}{4}
\providecommand{\BIBentryALTinterwordspacing}{\spaceskip=\fontdimen2\font plus
\BIBentryALTinterwordstretchfactor\fontdimen3\font minus
  \fontdimen4\font\relax}
\providecommand{\BIBforeignlanguage}[2]{{%
\expandafter\ifx\csname l@#1\endcsname\relax
\typeout{** WARNING: IEEEtran.bst: No hyphenation pattern has been}%
\typeout{** loaded for the language `#1'. Using the pattern for}%
\typeout{** the default language instead.}%
\else
\language=\csname l@#1\endcsname
\fi
#2}}
\providecommand{\BIBdecl}{\relax}
\BIBdecl

\bibitem{el2005real}
R.~El~Kaliouby and P.~Robinson, ``Real-time inference of complex mental states
  from facial expressions and head gestures,'' in \emph{Real-time vision for
  human-computer interaction}.\hskip 1em plus 0.5em minus 0.4em\relax Springer,
  2005, pp. 181--200.

\bibitem{baltruvsaitis2011real}
T.~Baltru{\v{s}}aitis, D.~McDuff, N.~Banda, M.~Mahmoud, R.~El~Kaliouby,
  P.~Robinson, and R.~Picard, ``Real-time inference of mental states from
  facial expressions and upper body gestures,'' in \emph{Automatic Face \&
  Gesture Recognition and Workshops (FG 2011), 2011 IEEE International
  Conference on}.\hskip 1em plus 0.5em minus 0.4em\relax IEEE, 2011, pp.
  909--914.

\bibitem{Morency2016}
T.~Baltrušaitis, P.~Robinson, and L.~P. Morency, ``Openface: An open source
  facial behavior analysis toolkit,'' in \emph{2016 IEEE Winter Conference on
  Applications of Computer Vision (WACV)}, March 2016, pp. 1--10.

\bibitem{cao2017realtime}
Z.~Cao, T.~Simon, S.-E. Wei, and Y.~Sheikh, ``Realtime multi-person 2d pose
  estimation using part affinity fields,'' in \emph{CVPR}, 2017.

\bibitem{simon2017hand}
T.~Simon, H.~Joo, I.~Matthews, and Y.~Sheikh, ``Hand keypoint detection in
  single images using multiview bootstrapping,'' in \emph{CVPR}, 2017.

\bibitem{wei2016cpm}
S.-E. Wei, V.~Ramakrishna, T.~Kanade, and Y.~Sheikh, ``Convolutional pose
  machines,'' in \emph{CVPR}, 2016.

\bibitem{Kanneman11}
D.~Kahneman, \emph{Thinking, fast and slow.}\hskip 1em plus 0.5em minus
  0.4em\relax Macmillan, 2011.

\bibitem{ekman1969nonverbal}
P.~Ekman and W.~V. Friesen, ``Nonverbal leakage and clues to deception,''
  \emph{Psychiatry}, vol.~32, no.~1, pp. 88--106, 1969.

\bibitem{poh2011advancements}
M.-Z. Poh, D.~J. McDuff, and R.~W. Picard, ``Advancements in noncontact,
  multiparameter physiological measurements using a webcam,'' \emph{IEEE
  transactions on biomedical engineering}, vol.~58, no.~1, pp. 7--11, 2011.

\bibitem{shotton2013real}
J.~Shotton, T.~Sharp, A.~Kipman, A.~Fitzgibbon, M.~Finocchio, A.~Blake,
  M.~Cook, and R.~Moore, ``Real-time human pose recognition in parts from
  single depth images,'' \emph{Communications of the ACM}, vol.~56, no.~1, pp.
  116--124, 2013.

\bibitem{stiefelhagen1997model}
R.~Stiefelhagen, J.~Yang, and A.~Waibel, ``A model-based gaze tracking
  system,'' \emph{International Journal on Artificial Intelligence Tools},
  vol.~6, no.~02, pp. 193--209, 1997.

\bibitem{charness2001perceptual}
N.~Charness, E.~M. Reingold, M.~Pomplun, and D.~M. Stampe, ``The perceptual
  aspect of skilled performance in chess: Evidence from eye movements,''
  \emph{Memory \& cognition}, vol.~29, no.~8, pp. 1146--1152, 2001.

\bibitem{Paletta2017}
\BIBentryALTinterwordspacing
L.~Paletta, A.~Dini, C.~Murko, S.~Yahyanejad, M.~Schwarz, G.~Lodron,
  S.~Ladst\"{a}tter, G.~Paar, and R.~Velik, ``Towards real-time probabilistic
  evaluation of situation awareness from human gaze in human-robot
  interaction,'' in \emph{Proceedings of the Companion of the 2017 ACM/IEEE
  International Conference on Human-Robot Interaction}, ser. HRI '17.\hskip 1em
  plus 0.5em minus 0.4em\relax New York, NY, USA: ACM, 2017, pp. 247--248.
  [Online]. Available: \url{http://doi.acm.org/10.1145/3029798.3038322}
\BIBentrySTDinterwordspacing

\bibitem{giraud2013multimodal}
T.~Giraud, M.~Soury, J.~Hua, A.~Delaborde, M.~Tahon, D.~A.~G. Jauregui,
  V.~Eyharabide, E.~Filaire, C.~Le~Scanff, L.~Devillers \emph{et~al.},
  ``Multimodal expressions of stress during a public speaking task: Collection,
  annotation and global analyses,'' in \emph{Affective Computing and
  Intelligent Interaction (ACII), 2013 Humaine Association Conference
  on}.\hskip 1em plus 0.5em minus 0.4em\relax IEEE, 2013, pp. 417--422.

\bibitem{abadi2013multimodal}
M.~K. Abadi, J.~Staiano, A.~Cappelletti, M.~Zancanaro, and N.~Sebe,
  ``Multimodal engagement classification for affective cinema,'' in
  \emph{Affective Computing and Intelligent Interaction (ACII), 2013 Humaine
  Association Conference on}.\hskip 1em plus 0.5em minus 0.4em\relax IEEE,
  2013, pp. 411--416.

\bibitem{reingold2005perception}
E.~M. Reingold and N.~Charness, ``Perception in chess: Evidence from eye
  movements,'' \emph{Cognitive processes in eye guidance}, pp. 325--354, 2005.

\bibitem{portaz:hal-01595775}
\BIBentryALTinterwordspacing
M.~Portaz, M.~Garcia, A.~Barbulescu, A.~Begault, L.~Boissieux, M.-P. Cani,
  R.~Ronfard, and D.~Vaufreydaz, ``{Figurines, a multimodal framework for
  tangible storytelling},'' in \emph{{WOCCI 2017 - 6th Workshop on Child
  Computer Interaction at ICMI 2017 - 19th ACM International Conference on
  Multi-modal Interaction}}, Glasgow, United Kingdom, Nov. 2017, author
  version. [Online]. Available: \url{https://hal.inria.fr/hal-01595775}
\BIBentrySTDinterwordspacing

\bibitem{vaufreydaz:hal-01095667}
\BIBentryALTinterwordspacing
D.~Vaufreydaz and A.~N{\`e}gre, ``{MobileRGBD, An Open Benchmark Corpus for
  mobile RGB-D Related Algorithms},'' in \emph{{13th International Conference
  on Control, Automation, Robotics and Vision}}, Singapour, Singapore, Dec.
  2014. [Online]. Available: \url{https://hal.inria.fr/hal-01095667}
\BIBentrySTDinterwordspacing

\bibitem{holmqvist2011eye}
K.~Holmqvist, M.~Nystr{\"o}m, R.~Andersson, R.~Dewhurst, H.~Jarodzka, and
  J.~Van~de Weijer, \emph{Eye tracking: A comprehensive guide to methods and
  measures}.\hskip 1em plus 0.5em minus 0.4em\relax OUP Oxford, 2011.

\bibitem{poole2006eye}
A.~Poole and L.~J. Ball, ``Eye tracking in hci and usability research,''
  \emph{Encyclopedia of human computer interaction}, vol.~1, pp. 211--219,
  2006.

\bibitem{ehmke2007identifying}
C.~Ehmke and S.~Wilson, ``Identifying web usability problems from eye-tracking
  data,'' in \emph{Proceedings of the 21st British HCI Group Annual Conference
  on People and Computers: HCI... but not as we know it-Volume 1}.\hskip 1em
  plus 0.5em minus 0.4em\relax British Computer Society, 2007, pp. 119--128.

\bibitem{facereader_presentation}
M.~Den~Uyl and H.~Van~Kuilenburg, ``The facereader: Online facial expression
  recognition,'' in \emph{Proceedings of measuring behavior}, vol.~30, 2005,
  pp. 589--590.

\bibitem{radboud}
O.~Langner, R.~Dotsch, G.~Bijlstra, D.~H. Wigboldus, S.~T. Hawk, and
  A.~Van~Knippenberg, ``Presentation and validation of the radboud faces
  database,'' \emph{Cognition and emotion}, vol.~24, no.~8, pp. 1377--1388,
  2010.

\bibitem{karolinska}
E.~Goeleven, R.~De~Raedt, L.~Leyman, and B.~Verschuere, ``The karolinska
  directed emotional faces: a validation study,'' \emph{Cognition and emotion},
  vol.~22, no.~6, pp. 1094--1118, 2008.

\bibitem{facereader_radboud}
G.~Bijlstra and R.~Dotsch, ``Facereader 4 emotion classification performance on
  images from the radboud faces database,'' 2015.

\bibitem{anzalone2015evaluating}
S.~M. Anzalone, S.~Boucenna, S.~Ivaldi, and M.~Chetouani, ``Evaluating the
  engagement with social robots,'' \emph{International Journal of Social
  Robotics}, vol.~7, no.~4, pp. 465--478, 2015.

\bibitem{harrigan1985self}
J.~A. Harrigan, ``Self-touching as an indicator of underlying affect and
  language processes,'' \emph{Social Science \& Medicine}, vol.~20, no.~11, pp.
  1161--1168, 1985.

\bibitem{johal2015cognitive}
W.~Johal, D.~Pellier, C.~Adam, H.~Fiorino, and S.~Pesty, ``A cognitive and
  affective architecture for social human-robot interaction,'' in
  \emph{Proceedings of the Tenth Annual ACM/IEEE International Conference on
  Human-Robot Interaction Extended Abstracts}.\hskip 1em plus 0.5em minus
  0.4em\relax ACM, 2015, pp. 71--72.

\bibitem{aigrain2016multimodal}
J.~Aigrain, M.~Spodenkiewicz, S.~Dubuisson, M.~Detyniecki, D.~Cohen, and
  M.~Chetouani, ``Multimodal stress detection from multiple assessments,''
  \emph{IEEE Transactions on Affective Computing}, 2016.

\end{thebibliography}

\end{document}